\documentclass{emulateapj}
\usepackage{graphicx,ifthen,url,float,color,amsmath,hyperref,cleveref,tabularx,array,ulem}
\newcommand{\ang}{$\mbox{\AA}$}
\newcommand{\um}{$\mu$m}
\newcommand{\ha}{\rm H$\alpha$}

\newcommand{\rpm}{\raisebox{.2ex}{$\scriptstyle\pm$}}
\newcommand{\vsigma}{v$_{\rm c}$/$\sigma$}

\newcommand{\kms}{\rm km\,s$^{-1}$}
\def\msol{\ifmmode{{\rm M}_{\odot}}\else{M$_{\odot}$}\fi}
\def\mstar{\ifmmode{{\rm M}_{\star}}\else{M$_{\star}$}\fi}
\def\vsigma{\ifmmode{{\rm v}_{\rm c}/\sigma}\else{v$_{\rm c}$/$\sigma$}\fi}

\newcommand{\hubunits}{{\rm km\,s$^{-1}$\,Mpc$^{-1}$}}

\def\ltsima{$\buildrel < \over \sim$}
\def\simlt{\lower.5ex\hbox{\ltsima}}
\def\gtsima{$\buildrel > \over \sim$}
\def\simgt{\lower.5ex\hbox{\gtsima}}

\newcolumntype{P}[1]{>{\centering\arraybackslash}p{#1}}
\newcolumntype{M}[1]{>{\centering\arraybackslash}m{#1}}

\shorttitle{\ha\ Flat Rotation Curve in $z=1.6$ Disk}
\shortauthors{P. M. Drew et al.}
\begin{document} 

\title{Evidence of a flat outer rotation curve in a starbursting disk galaxy at $z=1.6$}
\author{
Patrick M. Drew\altaffilmark{1} \href{https://orcid.org/0000-0003-3627-7485}{\includegraphics[scale=.08]{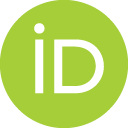}},
Caitlin M. Casey\altaffilmark{1} \href{https://orcid.org/0000-0002-0930-6466}{\includegraphics[scale=.08]{orcid_128x128.png}},
Anne D. Burnham\altaffilmark{1},
Chao-Ling Hung\altaffilmark{1,4},
Susan A. Kassin\altaffilmark{3,2} \href{https://orcid.org/0000-0002-3838-8093}{\includegraphics[scale=.08]{orcid_128x128.png}},
Raymond C. Simons\altaffilmark{2,3} \href{https://orcid.org/0000-0002-6386-7299}{\includegraphics[scale=.08]{orcid_128x128.png}},
and Jorge A. Zavala\altaffilmark{1} \href{https://orcid.org/0000-0002-7051-1100}{\includegraphics[scale=.08]{orcid_128x128.png}}}

\altaffiltext{1}{Department of Astronomy, The University of Texas at Austin, 2515 Speedway Blvd Stop C1400, Austin, TX 78712}
\email{Email: pdrew@utexas.edu}
\altaffiltext{2}{Johns Hopkins University, Baltimore, 3400 North Charles St., MD, 21218, USA}
\altaffiltext{3}{Space Telescope Science Institute, 3700 San Martin Dr., Baltimore, MD, 21218, USA}
\altaffiltext{4}{Department of Physics, Manhattan College, 4513 Manhattan College Pkwy Bronx, NY 10471}

\label{firstpage}

\begin{abstract}
Observations of the baryon to dark matter fraction in galaxies through cosmic time are a fundamental test for galaxy formation models.
Recent observational studies have suggested that some disk galaxies at $z>1$ host declining rotation curves, in contrast with observations of low redshift disk galaxies where stellar or HI rotation curves flatten at large radii.
We present an observational counterexample, a galaxy named DSFG850.95 at $z=1.555$ (4.1\,Gyr after the big bang) that hosts a flat rotation curve between radii of $\sim$6--14\,kpc (1.2--2.8 disk scale lengths) and has a dark matter fraction of $0.44\rpm0.08$ at the H-band half light radius, similar to the Milky Way.
We create position-velocity and position-dispersion diagrams using Keck/MOSFIRE spectroscopic observations of \ha\ and [NII] emission features, which reveal a flat rotation velocity of $V_{\rm flat}=285\rpm12$\,\kms\ and an 
ionized gas velocity dispersion of $\sigma_{0}=48\rpm4$\,\kms. This galaxy has a rotation-dominated velocity field with $V_{\rm flat}/\sigma_{0}\sim6$.
Ground-based H-band imaging reveals a disk with S\'ersic index of $1.29\rpm0.03$, an edge-on inclination angle of $87\rpm2^{\circ}$, and an H-band half light radius of $8.4\rpm0.1$\,kpc.
Our results point to DSFG850.95 being a massive, rotationally-supported disk galaxy with a high dark-matter-to-baryon fraction in the outer galaxy, similar to disk galaxies at low redshift.
\end{abstract}


\section{Introduction}\label{sec:introduction}
In the standard $\Lambda$CDM model of the Universe, the baryonic and dark matter components of galaxies grow in tandem from high to low redshift \citep[e.g.][]{White78a}. 
Galactic rotation curves are an important observational probe of these matter distributions in galaxies \citep[see][for a brief review]{Sofue01a}.
Through Keplerian inference, rotation curves provide a direct method for measuring the mass enclosed within a galactocentric radius. 
When combined with a light profile and a mass-to-light ratio, one may estimate the distribution of dark matter in a galaxy \citep[e.g.][]{Carignan85a,van-Albada85a,Kassin06a,de-Blok08a,Korsaga18a}.

At low redshift ($z$), disk galaxy rotation curves are seen to rise sharply in the inner few kpc and then flatten to a near-constant velocity \citep[e.g.][]{Rubin78a,McGaugh16a}. 
This is caused by baryons dominating the mass density in the inner regions of galaxies while dark matter dominates in the outer regions \citep{Kassin06a,Courteau07a,Dutton07a,Barnabe12a,Cappellari13a,Dutton13a,Martinsson13a,Martinsson13b,Courteau15a}.
Recent observational studies have begun to push galaxy rotation curve measurements to larger radii at $z$ \simgt{} 1 and while some find flat rotation curves \citep[e.g.][]{Di-Teodoro16a,Di-Teodoro18a,Xue18a}, others, intriguingly, find evidence of decreasing outer-galaxy rotation curves \citep{Lang17a,Genzel17a}. 
These studies attribute this to a combination of decreased dark matter in galaxies at intermediate redshifts as well as increased pressure support in the outer disks.
It is common for low redshift disk galaxies to show an initial decline before flattening in the outer disk \citep[e.g.][]{Rubin65a,Rubin78a,McGaugh16a}, but the low or negligible inferred dark-matter-to-total mass fractions in the \citeauthor{Lang17a} and \citeauthor{Genzel17a} samples imply that these initial declines will not begin to flatten if detected at higher radii.

Some studies find lower dark matter fractions in galaxies at $z$ \simgt{} 1 than at $z=0$ \citep{ForsterSchreiber09a,Price16a,Burkert16a,Wuyts16a} while others find no evolution \citep{Stott16a,Yuan17a,Di-Teodoro16a}. Many of these works rely on inverting scaling relations with significant scatter, such as the Kennicutt-Schmidt relation \citep{Kennicutt98a}, in order to estimate gas masses. This may be the source of disagreement.
There is much greater agreement among many of these studies when comparing the stellar to the total masses of the galaxies \citep[See also][]{Pelliccia17a}. 

Different simulations also make different predictions about the shapes of rotation curves at $z>1$. \citet{Teklu18a} find 38\% of disk galaxies with $\mstar>5\times10^{10}$\,\msol\ at $z=2$ in the cosmological hydrodynamic simulation, {\it Magneticum Pathfinder} (K. Dolag et al., in preparation) have declining rotation beyond $\sim$4\,kpc (as estimated from Figure 4 in \citealt{Teklu18a}).
In contrast, the IllustrisTNG cosmological hydrodynamic simulations \citep{Lovell18a} show initially declining rotation curves at $z=2$ that gradually flatten in the outer disk, similar to low-$z$ disk galaxies. It is clear that more observations of the outer rotation curves of high redshift galaxies and their dark matter fractions are needed to puzzle out the evolution of dark matter fractions with time. 

In this paper we present the serendipitously acquired rotation curve for a known dusty star forming galaxy at $z=1.555$, detected in \ha\ to a high galactocentric radius of 14\,kpc. The galaxy exhibits a flat rotation curve, well-fit by an arctangent profile and a dark matter fraction at 14\,kpc typical of disk galaxies at $z=0$.
The paper is organized as follows. Section \ref{sec:obs} details the observations and data reduction, Section \ref{sec:kinMod} discusses the kinematic modeling, Section \ref{sec:dynStel} details the dynamical and stellar mass estimations, and Section \ref{sec:summ} presents our conclusions.
Throughout this work we adopt a \textit{Planck} cosmology with H$_{0} = 67.7$\,\hubunits\ and $\Omega_{\Lambda} = 0.6911$ \citep{Planck-Collaboration16a}. We assume a Chabrier initial mass function (IMF; \citealt{Chabrier03a}).
\begin{figure}
    \centering
    \includegraphics[width=1.\columnwidth]{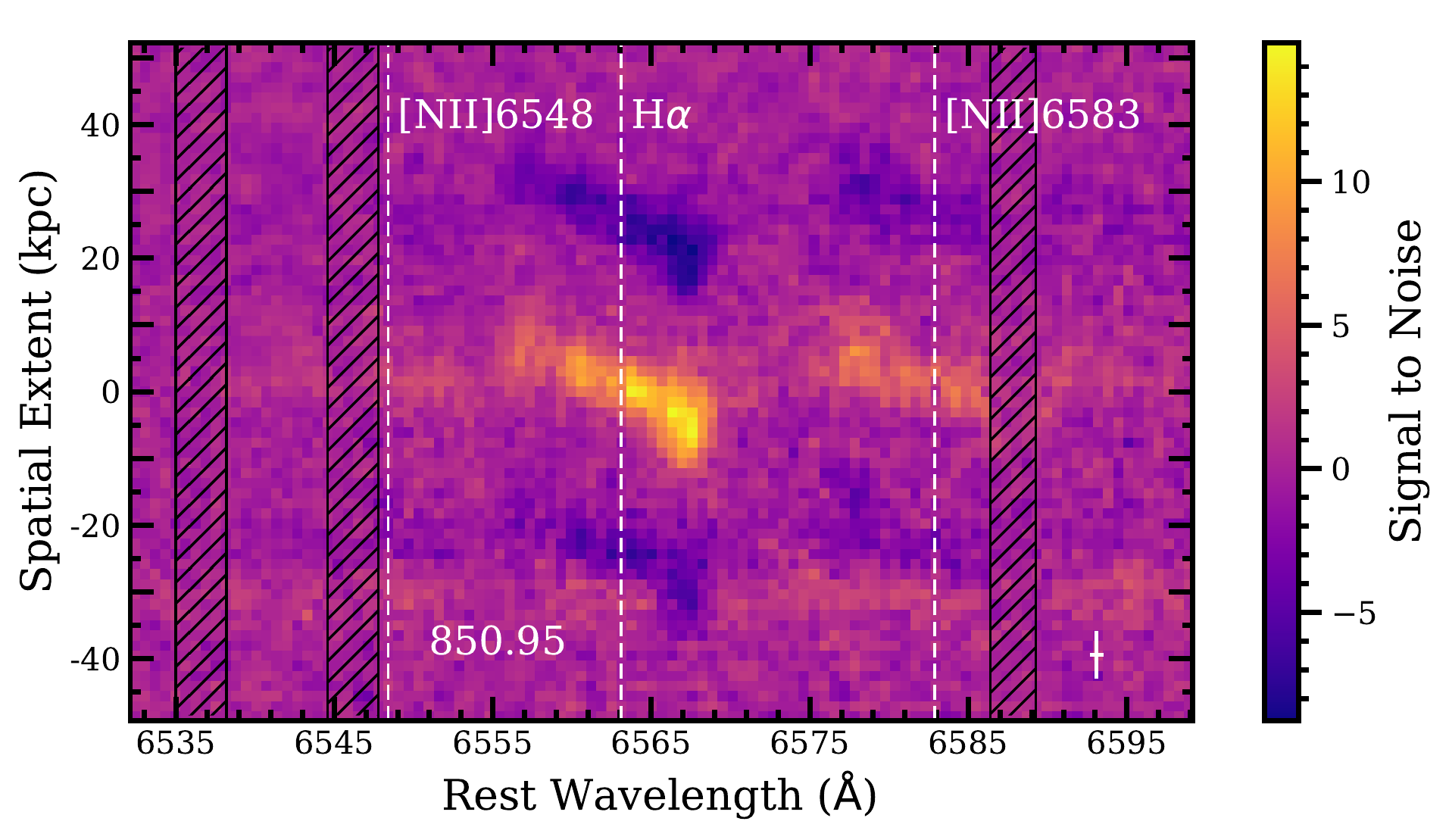}
    \caption{Two dimensional signal-to-noise spectrum of DSFG850.95 showing \ha, [NII], and continuum emission showing symmetric rotation with a flattened rotational velocity at high galactocentric radii. Hatched regions denote pixels significantly contaminated by telluric sky emission. Vertical dashed lines mark the location of \ha, [NII]$\lambda$6548, and [NII]$\lambda$6583.
    The white cross represents the resolution in each dimension. Negative reflections of the spectrum are caused by the standard ABBA nodding pattern used. 
    A separate foreground source at $z=0.7$ sits at a projected radius of $\sim$30\,kpc to the south of DSFG850.95 and is the source of continuum emission seen toward the bottom of the 2D spectrum.}
    \label{fig:snr}
\end{figure}

\section{Observations and Data Reduction}\label{sec:obs}
\begin{figure}
    \centering
    \includegraphics[width=\linewidth]{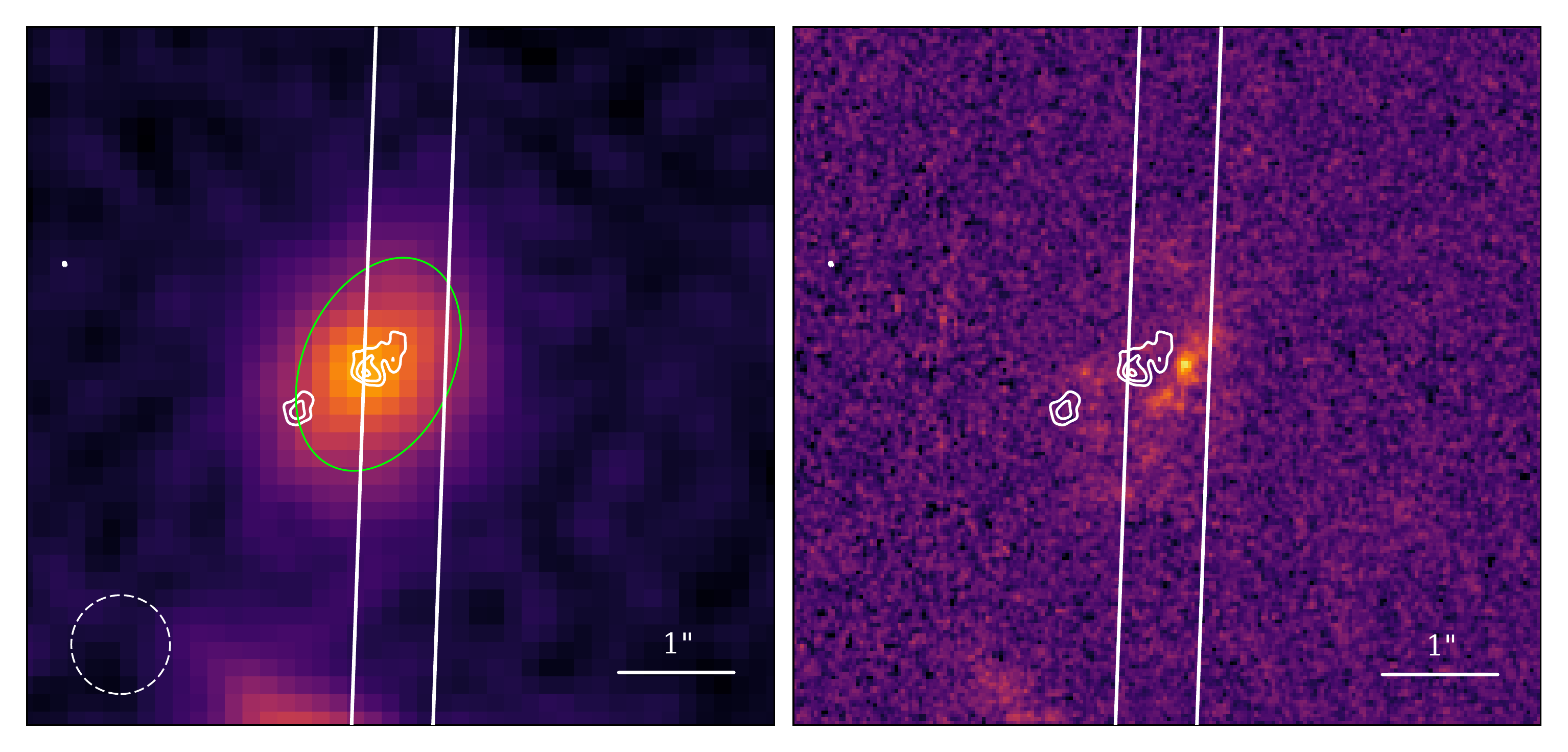}
    \caption{Imaging of DSFG850.95 in H band \citep[Left; rest-frame 6400\,\ang;][]{McCracken12a} and \textit{HST} F814W \citep[Right; rest-frame 3200\,\ang;][]{Koekemoer07a} with the MOSFIRE slit and ALMA dust contours at 3, 4, and 5$\sigma$ significance (Burnham et al. in preparation) in white. The white dashed circle shows the FWHM of seeing for the MOSFIRE observations.
    The green ellipse shows the H-band half-light radius.
    Emission seen to the south is from an unrelated foreground galaxy at a photometric redshift of 0.7.
    }\label{fig:image}
\end{figure}

The data were obtained with the Multi-Object Spectrometer For Infra-Red Exploration (MOSFIRE; \citealt{McLean10a,McLean12a}) on Keck I as part of an observational campaign to follow up a sample of 450\,\um- and 850\,\um-identified dusty star forming galaxies  \citep[DSFGs; for sample details see][]{Casey13a,Casey17a}. 
DSFGs comprise the most intense star-bursting galaxies in the Universe (See reviews by \citealt{Blain02a} and \citealt*{Casey14a}).

Observations were made on 2013 December 31 with the full width at half maximum (FWHM) of seeing between 0\farcs8--0\farcs9, as measured by the telescope focus routine, MIRA. We adopt an average seeing of $0\farcs85$ throughout this work. 
The galaxy, DSFG850.95, was observed in the H band for a total integration time of 1920\,s with the slit width set to $0\farcs7$. An 1\farcs5 ABBA nod pattern was used and the slit orientation was randomly placed with respect to our target due to position angle optimization driven by multiplexing.
Additional observational details are described for the full parent sample in \citet{Casey17a}.

The data were reduced with the MOSPY data reduction pipeline\footnote{\href{http://keck-datareductionpipelines.github.io/MosfireDRP/}{http://keck-datareductionpipelines.github.io/MosfireDRP/}}. 
\autoref{fig:snr} plots the reduced 2D signal to noise spectrum of DSFG850.95 showing \ha, [NII], and continuum emission. An unrelated foreground galaxy at photometric redshift of 0.7, COSMOS 2004241 \citep{Capak07a}, sits a projected $\sim$3\farcs5 to the south of DSFG850.95 and is seen in faint continuum emission in \autoref{fig:snr} at spatial extent $-30$\,kpc. It does not affect the analysis of DSFG850.95's rotation curve.

\begin{figure}
    \centering
    \includegraphics[width=\linewidth]{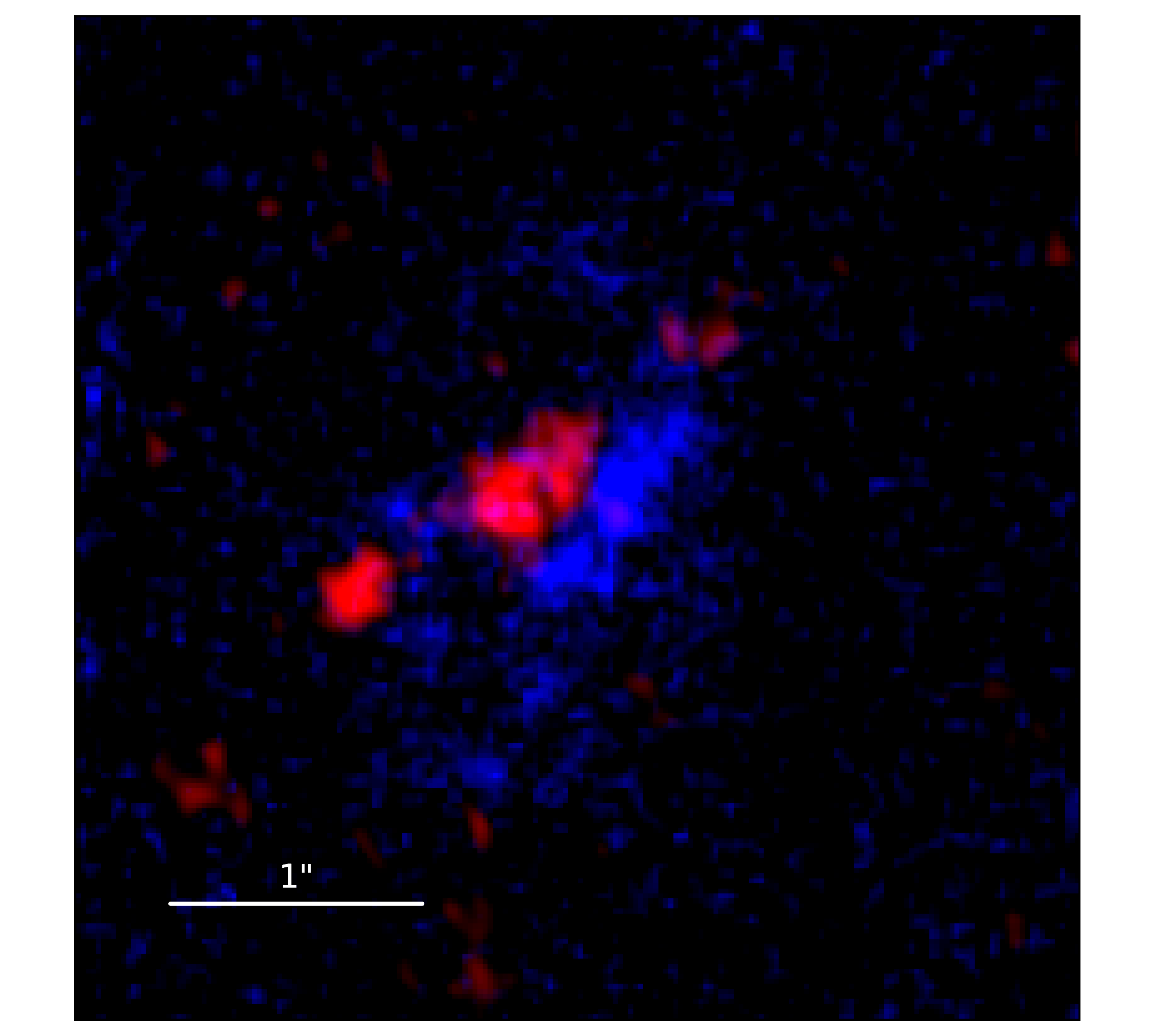}
    \caption{Imaging of DSFG850.95 in {\it HST} F814W (rest-frame UV) and ALMA 870\,\um\ dust continuum at $0\farcs1$ resolution in blue and red respectively.}\label{fig:rgb}
\end{figure}

The left panel of \autoref{fig:image} shows H-band imaging (rest-frame $\sim$6400\,\ang), MOSFIRE slit, and ALMA 870\,\um\ dust continuum contours of DSFG850.95. The H-band FWHM of seeing was comparable to that of the MOSFIRE observations and is shown as the dashed circle. The green ellipse shows the H-band half-light radius. There is a misalignment between the position angles of the H-band exponential disk fit and the slit of $23\rpm4^{\circ}$, which should not strongly affect our measured radial velocities. We will discuss this further in \autoref{subsec:lightprofile}.
The right panel of \autoref{fig:image} shows \textit{Hubble} imaging in the F814W filter, at a rest-frame effective wavelength of $\sim$3200\,\ang. 
The centroid of UltraVISTA K-band emission was used for the slit targeting position (rest-frame $\sim$8600\,\ang). It shares the same positional centroid as the H-band image in the left panel of \autoref{fig:image}, however the slit was slightly offset from the intended target position. The seeing on the night of the MOSFIRE observation is much larger than this offset, and is large enough to cause emission originating from positions outside the slit to be probed by the slit. This offset, therefore, does not strongly affect our analysis. Additionally, the rest-frame UV emission, which expected to originate in the same regions as the \ha\ emission is well-probed by the slit, though it is more highly extincted than the \ha\ emission.
The offset between the \textit{HST} and dust emission centroids does demonstrate the importance of dust obscuration in this dusty star forming galaxy, however.
\autoref{fig:rgb} shows another look at the \textit{HST} I-band imaging in blue with ALMA 870\,\um\ dust continuum in red.

We extracted 1D spectra from the MOSFIRE data using the {\sc iraf}\footnote{{\sc iraf} is distributed by the National Optical Astronomy Observatories, which are operated by the Association of Universities for Research in Astronomy, Inc., under cooperative agreement with the National Science Foundation} routine {\sc apall}. Apertures of extraction were placed on each pixel across the galaxy with a radius equal to half the average seeing. Spectra were extracted with variance weighting.
Radial velocities and velocity dispersions were measured from the extracted spectra using the {\sc splot} fitting routine in {\sc iraf}. 
We fit \ha, [NII]$\lambda 6548$, and [NII]$\lambda 6583$ simultaneously with fixed spacing between spectral features and tied but variable emission line full widths at half maximum (FWHM).
Pixels contaminated with telluric line emission were omitted from the fits and errors in the best-fit parameters were estimated using 1000 Monte Carlo perturbations of the data by typical sky noise. \autoref{fig:rotcurv} shows the position-velocity and position-dispersion diagrams measured from these fits. The velocity curve (upper panel), corrected for galaxy inclination angle (see \autoref{subsec:lightprofile} for details), rises steeply in the inner few kpc and then flattens to a near-constant velocity in the outer regions of the galaxy. The dispersion curve (lower panel) is centrally peaked due to the effect of beam smearing.

\section{Kinematic Model}\label{sec:kinMod}

We model the rotational velocity, $V(r)$, and velocity dispersion, $\sigma (r)$ of DSFG850.95 following the method employed by e.g. \citet{Courteau97a,Weiner06a}, modified to allow for the parameterization of a possible decline in rotational velocity at high galactocentric radii. The model assumes the intrinsic rotation curve follows a modified arctan function of the form:
\begin{gather}
    V(r) = \frac{2}{\pi}V_{\rm a}\arctan\left({\frac{r}{r_{\rm t}}}\right) +cr,
\end{gather}
where $V_{a}$ is the asymptotic velocity, $r_{t}$ is the knee radius, $r$ is the distance from the center of the galaxy, and $c$ is the outer-galaxy slope which allows us to quantify the degree to which the rotation curve increases, flattens, or decreases. 

We also measure an isotropic ionized gas velocity dispersion for the entire galaxy, $\sigma_0$, which is taken to be the vertical offset of a Gaussian fitted to the position-dispersion data in the bottom panel of \autoref{fig:rotcurv}:
\begin{gather}
    \Sigma(r) = A\exp\left(-\frac{(r-m)^{2}}{2w^{2}}\right)+\sigma_{0},
\end{gather}
where $\Sigma(r)$ is the value of the fitted Gaussian at each $r$, $A$ is the amplitude, $r$ is the distance from the center of the galaxy, $m$ is the central position, $w$ is the width of the Gaussian, and $\sigma_{0}$ is the the isotropic velocity dispersion. 
We choose $\sigma_{0}$ as the representative dispersion for DSFG850.95 because it is the value the velocity dispersion approaches in the outer galaxy where beam smearing has less of an effect.
Beam smearing is well known to artificially increase the observed velocity dispersion toward the centers of galaxies due to the rise in the rotation curve
\citep[e.g.][]{Bosma78a,Weiner06a,Di-Teodoro15b}, but has less of an effect at higher galactocentric radii.

\begin{table}[]
    \caption{{\sc Physical Characteristics of DSFG850.95}}
    \centering
    \begin{tabularx}{\columnwidth}{M{4cm}|M{4cm}}
        \hline\hline
        {\sc Property} & {\sc Value} \\
        \hline
        $z$ & 1.555 \\
        $V_{\rm flat}$\footnote{Average of flat rotation on both sides of DSFG850.95.} & $285\rpm12$\,\kms\\
        $\sigma_{0}$ & $48\rpm4$\,\kms \\
        $V_{\rm flat}/\sigma_{0}$ & 6 \\
        $\Delta$P.A.\footnote{Position angle offset between slit and H-band semi-major axis.} & $23\rpm4^{\circ}$\\
        n\footnote{S\'ersic Index.} & $1.29\rpm0.03$ \\
        $i$\footnote{Inclination angle.} & $87\rpm2^{\circ}$ \\
        $r_{1/2}$ & $8.4\rpm0.1$\,kpc \\ 
        $r_{2.2}$ & $11.0\rpm0.2$\,kpc \\
        $r_{\rm max}$\footnote{Maximum detected radius of \ha.} & $14.1$\,kpc \\
        SFR & $280^{+110}_{-90}$\,\msol{}\,yr$^{-1}$ \\
        M$_{\star}$ & ($3.8\rpm3.0$)$\times10^{10}$\,\msol{} \\
        M$_{\rm H_{2}}$ & ($8.88\rpm 0.03$)$\times10^{10}\,\msol{}$ \\
        M$_{\rm dyn}$(14\,kpc) & (2.7$\rpm$0.3)$\times10^{11}$\,\msol{} \\
        M$_{\rm dyn}$(R$_{1/2}$) & (1.9$\rpm$0.1)$\times10^{11}$\,\msol{} \\
        $f_{DM}$(R$_{1/2}$) & $0.44\rpm0.08$ \\
        \hline\hline
        
    \end{tabularx}
    \label{tab:properties}
\end{table}

We correct the observed velocity dispersions for the intrinsic instrumental dispersion prior to fitting using the equation: $\sigma=\sqrt{\sigma^{2}_{obs}-\sigma^{2}_{inst}}$
where $\sigma$ is the velocity dispersion reported in \autoref{fig:rotcurv}, $\sigma_{obs}$ is the FWHM of the emission line divided by 2.355 and $\sigma_{inst}$ is the combined instrumental and spectral seeing dispersion, taken to be the average width of several telluric lines measured across the unreduced spectrum ($\sigma_{inst}=39\rpm2$\,\kms). The reported error on $\sigma_{inst}$ is the standard deviation in the mean of the measured telluric line widths and are folded into the errors in velocity dispersion in \autoref{fig:rotcurv}. Measured velocity dispersions that are less than the instrumental dispersion are unreliable because small errors on the observed dispersion correspond to large changes in the inferred physical dispersion \citep{Weiner06a}. Dispersions measured below the instrumental seeing are plotted as upper limits at 2$\sigma$ significance in the bottom panel of \autoref{fig:rotcurv} and are excluded from the dispersion curve fits, though their central wavelengths are still fitted for the velocity curve.

We fit the position-velocity and position-dispersion data with the pymc3 python package for Markov Chain Monte Carlo (MCMC; \citealt{Salvatier16a}) so that we may characterize the posterior probability distributions of our fit parameters. 
The light red curves in the top panel of \autoref{fig:rotcurv} show fifty randomly-chosen MC fits and the best-fit curve is plotted in dark red. The best-fit value of asymptotic velocity is $|V_{a}|=310\rpm6$\,\kms, the knee radius is $r_{t} = 0\farcs17\rpm0\farcs02$ or $1.5\rpm 0.2$\,kpc, and the outer slope is $c=-0.5\rpm1.0$\,\kms\,arcsec$^{-1}$, which is consistent with a flat slope. The best-fit velocity curve is still seen to be rising at the radius of the outermost datapoint. There is no evidence of a velocity turnover with declining rotation. The average velocity of the flat portion, taken to be the average of the datapoints between radii of $\sim$6--14\,kpc, corresponding to 1.2--2.8 disk scale lengths, is $V_{\rm flat}=285\rpm12$\,\kms.

The bottom panel of \autoref{fig:rotcurv} shows the measured velocity dispersion data with fifty randomly-selected MC fits overplotted in grey with the best-fit curve in black. The best-fit dispersion, $\sigma_0$, is 48$\rpm$4\,\kms, typical of mildly turbulent disk galaxies at $z\sim1.5$ \citep[e.g.][]{Wright07b,Law09a,Epinat12a,Contini12a,Wisnioski15a,Simons17a}.

\subsection{H-Band Light Profile Fitting}\label{subsec:lightprofile}

\begin{figure}
    \centering
    \includegraphics[width=\linewidth]{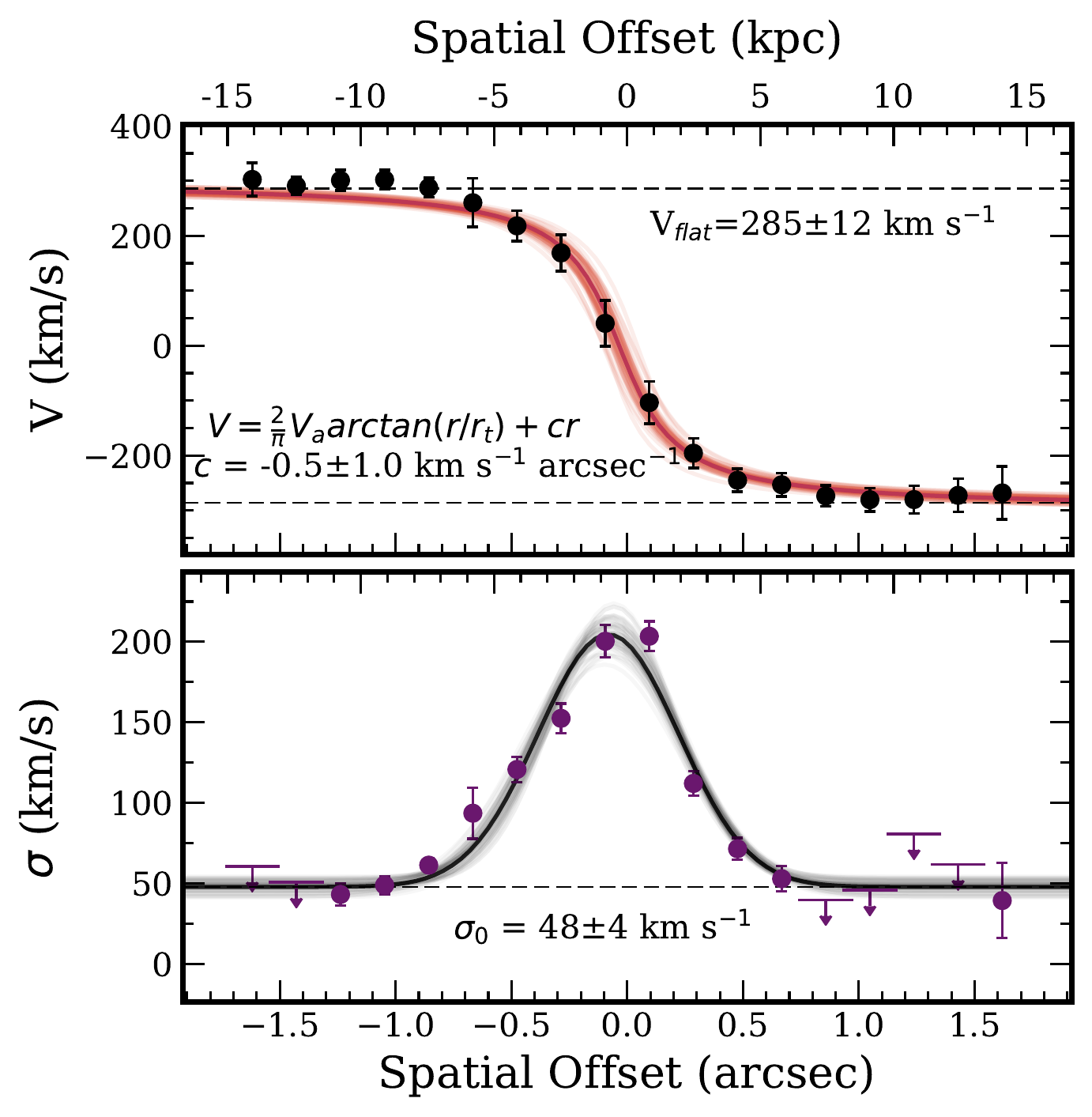}
    \caption{Position-velocity and position-dispersion diagrams of DSFG850.95 showing flattened rotation between $\sim$6--14\,kpc with average velocity of $V_{\rm flat}=285\rpm12$\,\kms, marked as the dashed lines in the upper panel. The light red curves show 50 randomly-selected Monte Carlo chains from the velocity curve fit with the dark red curve showing the best fit. The outer-galaxy slope of $c=-0.5\rpm1.0$\,\kms\,arcsec$^{-1}$ is consistent with flat rotation.
    The grey curves show 50 randomly-selected Monte Carlo chains from the dispersion fit and the black curve shows the best fit with galaxy-wide ionized gas velocity dispersion of $\sigma_{0}=48\rpm4$\,\kms, marked as the dashed line in the lower panel, implying DSFG850.95 is rotation dominated ($V_{\rm flat}/\sigma_{0}\approx6$).
    Rotational velocities are measured from spectral extraction apertures centered on each pixel in the 2D spectrum with a radius of half the average seeing and are corrected for a galaxy inclination of $i=87\rpm2^{\circ}$. Upper limits on the velocity dispersion are at 2$\sigma$ significance where the measured dispersion is less than the combined instrumental and seeing dispersion as measured from telluric line widths.}
    \label{fig:rotcurv}
\end{figure}

Using the 2D image fitting code, GALFIT \citep{Peng02a}, we fit a S\'ersic profile \citep{Sersic63a} to H-band imaging of DSFG850.95 from the UltraVISTA survey \citep[][see \autoref{fig:image}, above]{McCracken12a}. 

The fit reveals a S\'ersic index of $n=1.29\rpm0.03$, implying an exponential disk light profile. Adding a second S\'ersic component to fit a bulge causes significantly worse fits. The H-band half-light radius is $r_{1/2}=8.4\rpm0.1$\,kpc, or using the relation $r_d=r_{1/2}/1.68$, the disk scale radius is $r_d = 5.0\rpm0.1$\,kpc and $r_{2.2}=11.0\rpm0.2$\,kpc.
We divide the velocities in the top panel of \autoref{fig:rotcurv} by $\sin (i)$ to correct for galaxy inclination, assuming an intrinsic edge-on axis ratio of 0.25 \citep{Wuyts16a}. From our best-fit axis ratio, $b/a=0.668\rpm0.008$, we determine the galaxy has an approximately edge-on inclination of $i=87\rpm 2^{\circ}$ using the equation \citep[e.g.][]{Wuyts16a}:
\begin{gather}
    \cos{i} = \sqrt{\frac{(b/a)^2-0.25^2}{1-0.25^2}}
\end{gather}
\autoref{tab:properties} summarizes the physical characteristics of DSFG850.95 from GALFIT and our other analyses.

There is a misalignment of $23\rpm4^{\circ}$ between the position angles of the H-band semi-major axis and the slit.
Large misalignments would result in an underprediction of the maximum rotational velocity of a galaxy, however \citet{Weiner06a} shows that the average decrease in observed rotational velocities for a slit misalignment of $\sim23^{\circ}$ is small --- approximately the same as the error on our measurement of $V_{\rm flat}$.

The H-band imaging, together with the rotational velocity and velocity dispersion analyses in \autoref{sec:kinMod} suggest DSFG850.95 is a rotationally-dominated disk galaxy at $z=1.555$ that is similar to massive disk galaxies at low redshift.

\section{Dynamical, Stellar, Gas, and Dark Matter Masses}\label{sec:dynStel}

Our radial velocity measurements as a function of radius, corrected for galaxy inclination and ionized gas velocity dispersion support allow us to measure the total mass enclosed as a function of radius in the galaxy using the equation
\citep[e.g.][]{Erb03a,Swinbank04a,van-Starkenburg08a,Epinat09a,Di-Teodoro18a}:
\begin{gather}\label{eq:mdyn}
    M_{dyn}(r) = \frac{V_{c}^{2}(r)r}{G},
\end{gather}
where $G$ is the gravitational constant, $r$ is the galactocentric radius, and $V_{c}$ is the circular velocity, given by:
\begin{gather}
V_{c}=\sqrt{V^2+2\sigma_{0}^2\frac{r}{r_d}},
\end{gather}
where $V$ is the inclination-corrected rotational velocity, $\sigma_{0}$ (measured to be 48$\rpm$4\,\kms) is the isotropic ionized gas velocity dispersion, and $r_d$ is the exponential disk scale length measured from H-band imaging.
The median difference between the rotational and circular velocities is 2\%, since $\sigma_{0}$ is small compared to $V$.

\begin{figure}
    \centering
    \includegraphics[width=0.99\columnwidth]{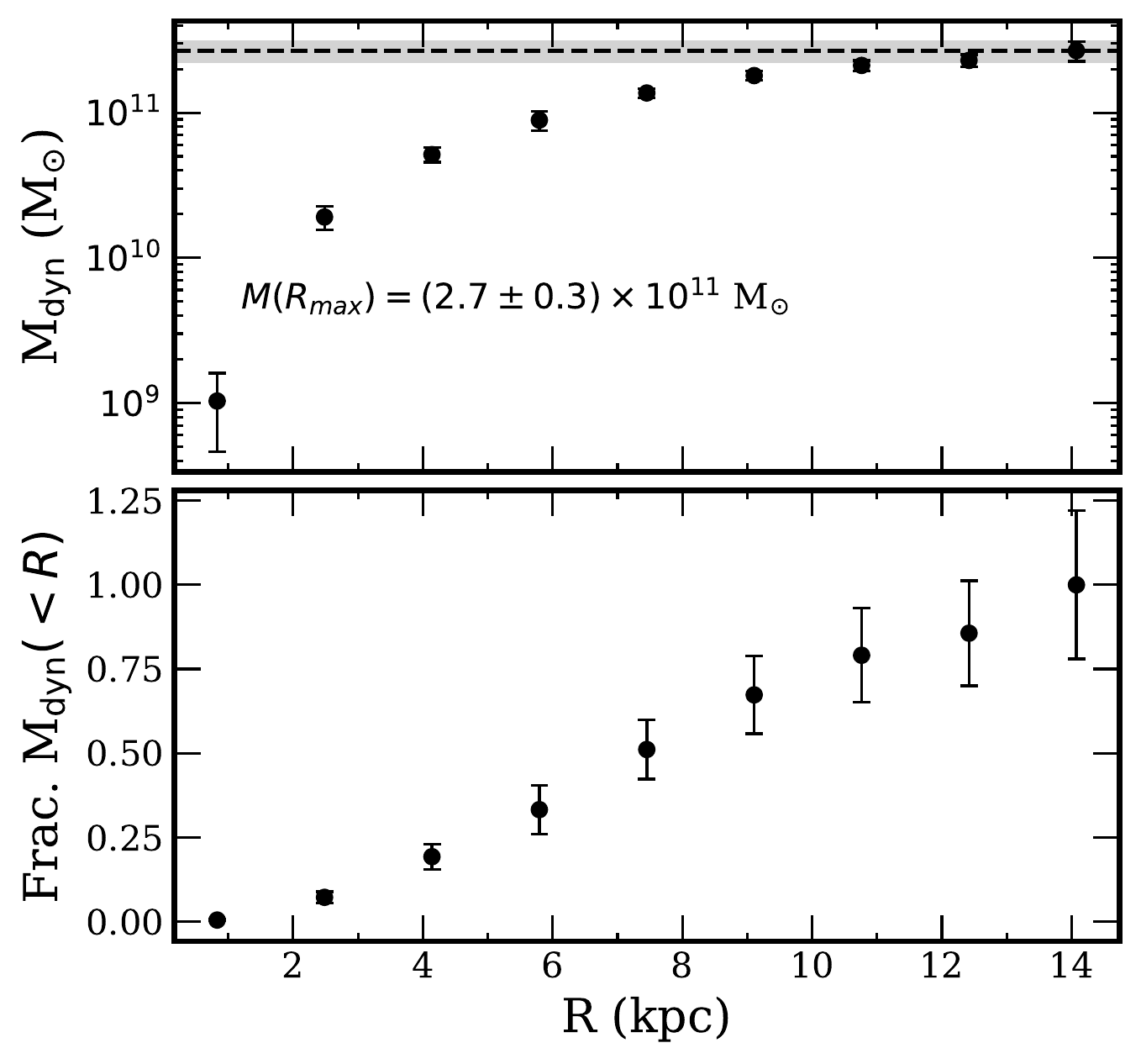}
    \caption{Top: The dynamical mass as a function of radius using $M_{dyn}=V_{c}^{2}(R)R/G$, where $V_{c}$ is the circular velocity, $R$ is radius, and $G$ is the gravitational constant. The horizontal black dotted line is the dynamical mass at the largest radius, $\sim$14\,kpc, and the grey shaded region denotes the extent of the errors on that measurement. Datapoints are the average from both sides of the galaxy. Bottom: The fraction of dynamical mass enclosed as a function of galactocentric radius.}
    \label{fig:mDyn}
\end{figure}

The top panel of \autoref{fig:mDyn} shows the dynamical mass as a function of radius.
The total dynamical mass enclosed within the H-band half light radius is $(1.9\rpm0.1)\times 10^{11}$\,\msol{}, and within 14\,kpc
is $(2.7\rpm0.3) \times 10^{11}$\,\msol{}. The horizontal dashed line and shaded region in the top panel of \autoref{fig:mDyn} marks the maximum dynamical mass and its associated error. The bottom panel of \autoref{fig:mDyn} shows the fraction of dynamical mass enclosed by radius.

Next we compare the dynamical mass at the half light radius with stellar and gas masses.
\citet{Casey17a} uses MAGPHYS with the high-$z$ extension \citep{da-Cunha08a,da-Cunha15a} 
to fit UV through sub-mm ancillary photometry from the COSMOS collaboration \citep{Capak07a,Laigle16a} 
and finds a stellar mass of $\mstar=(3.8^{+0.4}_{-1.0})\times10^{10}$\,\msol{}. 
We follow the procedure of \citet{Hainline11a} for similarly-selected DSFGs at similar redshifts to derive a more conservative error than \citeauthor{Casey17a} 
We take the error to be half the difference between the stellar mass estimated with an instantaneous burst history and that estimated with a continuous star formation history. Thus, we arrive at a total stellar mass of $\mstar{}=(3.8\rpm3.0)\times 10^{10}$\,\msol{}.
\autoref{tab:properties} summarizes all the estimated and measured parameters of this galaxy.

\begin{figure}
    \centering
    \includegraphics[width=0.99\columnwidth]{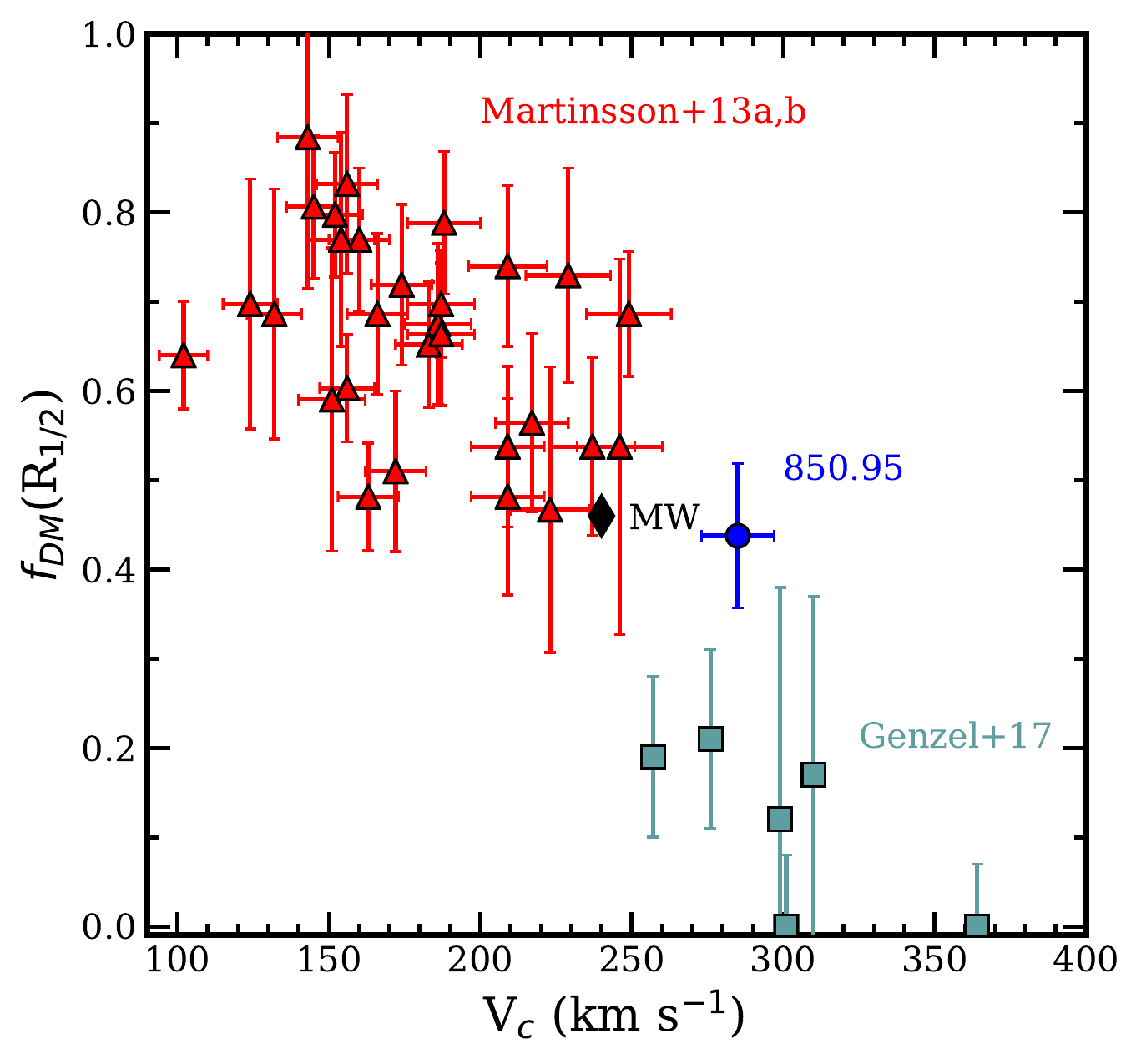}
    \caption{The fraction of dark matter at the half light radius as a function of circular velocity at the half light radius for different samples in the literature. The red triangles show $z=0$ disk galaxies \citep{Martinsson13a,Martinsson13b}, the Milky Way is plotted as the black diamond (\citealt{BlandHawthorn16a}), and the \citet{Genzel17a} sample is plotted as light blue squares. The dark matter fraction of DSFG850.95 is more similar to that of the Milky Way, albeit at a higher circular velocity.}
    \label{fig:dmFrac}
\end{figure}

Following the prescription of \citet{Scoville16a} we use our ALMA 870\,\um\ dust continuum observations to estimate a molecular gas mass of $(8.9\rpm 0.4)\times 10^{10}$\,\msol{}, all of which is contained within the half light radius.
We calculate the dark matter fraction at the half light radius using the equation:
\begin{gather}
    f_{\rm DM}(R_{1/2}) = 1-\frac{M_{\star}(R_{1/2})+M_{gas}}{M_{dyn}(R_{1/2})},
\end{gather}
where $M_{\star}(R_{1/2})$ is half of the total stellar mass, $M_{\rm gas}$ is the gas mass, and $M_{dyn}(R_{1/2})$ is the dynamical mass at the half light radius. We find $f_{\rm DM}(R_{1/2})=0.44\rpm0.08$. \autoref{fig:dmFrac} shows the dark matter fraction as a function of circular velocities for a few studies in the literature \citep{Martinsson13a,Martinsson13b,BlandHawthorn16a,Genzel17a}. DSFG850.95 has a dark matter fraction closer to that of the Milky Way than the galaxies in \citet{Genzel17a}, though it has a higher rotational velocity than the Milky Way.


We conclude by noting that our finding of a flat rotation curve may not be inconsistent with the conclusions of \citet{Lang17a}. The majority of disk galaxies in cosmological hydrodynamic simulations by \citet{Teklu18a} exhibit flat rotation curves with only 38\% hosting declining rotation curves.
\citet{Lang17a} split their data out into two subsamples according to the degree of pressure support, one with $V_{rot}/\sigma_{0}>6.3$ and the other $<6.3$. They define $V_{rot}$ as the intrinsic, inclination corrected peak rotation velocity and $\sigma_{0}$ as the intrinsic velocity dispersion, both equivalent to our $V_{c}$ and $\sigma_{0}$. The galaxies with less pressure support ($V_{rot}/\sigma_{0}>6.3$) have flatter outer rotation curves than those with more pressure support, supporting the conclusion that much of the observed velocity decrease in their sample is due to increased pressure support. Our rotation curve, with a $V_{c}/\sigma_{0}\sim 6$ is consistent with their stacked curve of galaxies with $V_{rot}/\sigma_{0}>6.3$. A statistical sample of galaxy rotation curves detected well beyond 10\,kpc is needed to determine the fraction of flat or declining rotation curves at high redshift.

\section{Summary}\label{sec:summ}
Recent observations of disk galaxy rotation curves have found declining rotational velocities beyond $\approx$2.2 exponential disk scale lengths, implying intermediate redshift disk galaxies have lower dark matter fractions than $z=0$ galaxies as well as increased velocity dispersion support.

In this paper we present a clear counterexample from MOSFIRE observations of \ha\ and [NII] emission in the $z=1.555$ disk galaxy, DSFG850.95, showing a flat rotation curve between $\sim$6--14\,kpc (1.2--2.8 disk scale lengths), well-fit by an arctangent function. At the H-band half light radius we estimate the dark matter fraction to be $0.44\rpm 0.08$ based on UV-through-sub-mm photometry and 870\,\um\ dust continuum observations. Ground-based H-band imaging reveals a sersic index of $n=1.29\rpm 0.03$ and an edge-on inclination angle. Taken together, these data suggest DSFG850.95 is a massive, rotationally supported disk galaxy in place by $z=1.555$ (4.1\,Gyr after the big bang) with a flat rotation curve and a dark matter fraction at the half light radius, similar to low redshift disk galaxies.

\acknowledgements
The authors thank the anonymous referee for a
constructive report that significantly strengthened the arguments presented in this paper. The authors also wish to thank Michael Boylan-Kolchin, Michael Petersen, and Justin Spilker for useful discussions. P.M.D. and C.M.C. acknowledge financial support from NASA Keck PI Data Awards: 2012B-U039M, 2011B-H251M, 2012B-N114M, 2017A-N136M, the University of Texas at Austin College of Natural Sciences, and NSF grants AST-1714528 and AST-1814034. This paper makes use of ALMA data ADS/JAO.ALMA\#2015.1.00568S. ALMA is a partnership of ESO (representing its member states), NSF (USA) and NINS (Japan), together with NRC (Canada) and NSC and ASIAA (Taiwan) and KASI (Republic of Korea), in cooperation with the Republic of Chile. The Joint ALMA Observatory is operated by ESO, AUI/NRAO and NAOJ. The authors wish to recognize and acknowledge the very significant cultural role and reverence that the summit of Mauna Kea has always had within the indigenous Hawaiian community. We are most fortunate to have the opportunity to conduct observations from this mountain.


\begin{thebibliography}{}
\expandafter\ifx\csname natexlab\endcsname\relax\def\natexlab#1{#1}\fi

\bibitem[{{Barnab{\`e}} {et~al.}(2012){Barnab{\`e}}, {Dutton}, {Marshall},
  {Auger}, {Brewer}, {Treu}, {Bolton}, {Koo}, \& {Koopmans}}]{Barnabe12a}
{Barnab{\`e}}, M., {Dutton}, A.~A., {Marshall}, P.~J., {et~al.} 2012, \mnras,
  423, 1073

\bibitem[{{Blain} {et~al.}(2002){Blain}, {Smail}, {Ivison}, {Kneib}, \&
  {Frayer}}]{Blain02a}
{Blain}, A.~W., {Smail}, I., {Ivison}, R.~J., {Kneib}, J.-P., \& {Frayer},
  D.~T. 2002, \physrep, 369, 111

\bibitem[{{Bland-Hawthorn} \& {Gerhard}(2016)}]{BlandHawthorn16a}
{Bland-Hawthorn}, J., \& {Gerhard}, O. 2016, \araa, 54, 529

\bibitem[{{Bosma}(1978)}]{Bosma78a}
{Bosma}, A. 1978, PhD thesis, PhD Thesis, Groningen Univ., (1978)

\bibitem[{{Burkert} {et~al.}(2016){Burkert}, {F{\"o}rster Schreiber}, {Genzel},
  {Lang}, {Tacconi}, {Wisnioski}, {Wuyts}, {Bandara}, {Beifiori}, {Bender},
  {Brammer}, {Chan}, {Davies}, {Dekel}, {Fabricius}, {Fossati}, {Kulkarni},
  {Lutz}, {Mendel}, {Momcheva}, {Nelson}, {Naab}, {Renzini}, {Saglia},
  {Sharples}, {Sternberg}, {Wilman}, \& {Wuyts}}]{Burkert16a}
{Burkert}, A., {F{\"o}rster Schreiber}, N.~M., {Genzel}, R., {et~al.} 2016,
  \apj, 826, 214

\bibitem[{{Capak} {et~al.}(2007){Capak}, {Aussel}, {Ajiki}, {McCracken},
  {Mobasher}, {Scoville}, {Shopbell}, {Taniguchi}, {Thompson}, {Tribiano},
  {Sasaki}, {Blain}, {Brusa}, {Carilli}, {Comastri}, {Carollo}, {Cassata},
  {Colbert}, {Ellis}, {Elvis}, {Giavalisco}, {Green}, {Guzzo}, {Hasinger},
  {Ilbert}, {Impey}, {Jahnke}, {Kartaltepe}, {Kneib}, {Koda}, {Koekemoer},
  {Komiyama}, {Leauthaud}, {Le Fevre}, {Lilly}, {Liu}, {Massey}, {Miyazaki},
  {Murayama}, {Nagao}, {Peacock}, {Pickles}, {Porciani}, {Renzini}, {Rhodes},
  {Rich}, {Salvato}, {Sanders}, {Scarlata}, {Schiminovich}, {Schinnerer},
  {Scodeggio}, {Sheth}, {Shioya}, {Tasca}, {Taylor}, {Yan}, \&
  {Zamorani}}]{Capak07a}
{Capak}, P., {Aussel}, H., {Ajiki}, M., {et~al.} 2007, \apjs, 172, 99

\bibitem[{{Cappellari} {et~al.}(2013){Cappellari}, {McDermid}, {Alatalo},
  {Blitz}, {Bois}, {Bournaud}, {Bureau}, {Crocker}, {Davies}, {Davis}, {de
  Zeeuw}, {Duc}, {Emsellem}, {Khochfar}, {Krajnovi{\'c}}, {Kuntschner},
  {Morganti}, {Naab}, {Oosterloo}, {Sarzi}, {Scott}, {Serra}, {Weijmans}, \&
  {Young}}]{Cappellari13a}
{Cappellari}, M., {McDermid}, R.~M., {Alatalo}, K., {et~al.} 2013, \mnras, 432,
  1862

\bibitem[{{Carignan} \& {Freeman}(1985)}]{Carignan85a}
{Carignan}, C., \& {Freeman}, K.~C. 1985, \apj, 294, 494

\bibitem[{{Casey} {et~al.}(2014){Casey}, {Narayanan}, \& {Cooray}}]{Casey14a}
{Casey}, C.~M., {Narayanan}, D., \& {Cooray}, A. 2014, \physrep, 541, 45

\bibitem[{{Casey} {et~al.}(2013){Casey}, {Chen}, {Cowie}, {Barger}, {Capak},
  {Ilbert}, {Koss}, {Lee}, {Le Floc'h}, {Sanders}, \& {Williams}}]{Casey13a}
{Casey}, C.~M., {Chen}, C.-C., {Cowie}, L.~L., {et~al.} 2013, \mnras, 436, 1919

\bibitem[{{Casey} {et~al.}(2017){Casey}, {Cooray}, {Killi}, {Capak}, {Chen},
  {Hung}, {Kartaltepe}, {Sanders}, \& {Scoville}}]{Casey17a}
{Casey}, C.~M., {Cooray}, A., {Killi}, M., {et~al.} 2017, \apj, 840, 101

\bibitem[{{Chabrier}(2003)}]{Chabrier03a}
{Chabrier}, G. 2003, \pasp, 115, 763

\bibitem[{{Contini} {et~al.}(2012){Contini}, {Garilli}, {Le F{\`e}vre},
  {Kissler-Patig}, {Amram}, {Epinat}, {Moultaka}, {Paioro}, {Queyrel}, {Tasca},
  {Tresse}, {Vergani}, {L{\'o}pez-Sanjuan}, \& {Perez-Montero}}]{Contini12a}
{Contini}, T., {Garilli}, B., {Le F{\`e}vre}, O., {et~al.} 2012, \aap, 539, A91

\bibitem[{{Courteau}(1997)}]{Courteau97a}
{Courteau}, S. 1997, \aj, 114, 2402

\bibitem[{{Courteau} \& {Dutton}(2015)}]{Courteau15a}
{Courteau}, S., \& {Dutton}, A.~A. 2015, \apjl, 801, L20

\bibitem[{{Courteau} {et~al.}(2007){Courteau}, {Dutton}, {van den Bosch},
  {MacArthur}, {Dekel}, {McIntosh}, \& {Dale}}]{Courteau07a}
{Courteau}, S., {Dutton}, A.~A., {van den Bosch}, F.~C., {et~al.} 2007, \apj,
  671, 203

\bibitem[{{da Cunha} {et~al.}(2008){da Cunha}, {Charlot}, \&
  {Elbaz}}]{da-Cunha08a}
{da Cunha}, E., {Charlot}, S., \& {Elbaz}, D. 2008, \mnras, 388, 1595

\bibitem[{{da Cunha} {et~al.}(2015){da Cunha}, {Walter}, {Smail}, {Swinbank},
  {Simpson}, {Decarli}, {Hodge}, {Weiss}, {van der Werf}, {Bertoldi},
  {Chapman}, {Cox}, {Danielson}, {Dannerbauer}, {Greve}, {Ivison}, {Karim}, \&
  {Thomson}}]{da-Cunha15a}
{da Cunha}, E., {Walter}, F., {Smail}, I.~R., {et~al.} 2015, \apj, 806, 110

\bibitem[{{de Blok} {et~al.}(2008){de Blok}, {Walter}, {Brinks},
  {Trachternach}, {Oh}, \& {Kennicutt}}]{de-Blok08a}
{de Blok}, W.~J.~G., {Walter}, F., {Brinks}, E., {et~al.} 2008, \aj, 136, 2648

\bibitem[{{Di Teodoro} \& {Fraternali}(2015)}]{Di-Teodoro15b}
{Di Teodoro}, E.~M., \& {Fraternali}, F. 2015, \mnras, 451, 3021

\bibitem[{{Di Teodoro} {et~al.}(2016){Di Teodoro}, {Fraternali}, \&
  {Miller}}]{Di-Teodoro16a}
{Di Teodoro}, E.~M., {Fraternali}, F., \& {Miller}, S.~H. 2016, \aap, 594, A77

\bibitem[{{Di Teodoro} {et~al.}(2018){Di Teodoro}, {Grillo}, {Fraternali},
  {Gobat}, {Karman}, {Mercurio}, {Rosati}, {Balestra}, {Caminha}, {Caputi},
  {Lombardi}, {Suyu}, {Treu}, \& {Vanzella}}]{Di-Teodoro18a}
{Di Teodoro}, E.~M., {Grillo}, C., {Fraternali}, F., {et~al.} 2018, \mnras,
  arXiv:1801.06546

\bibitem[{{Dutton} {et~al.}(2007){Dutton}, {van den Bosch}, {Dekel}, \&
  {Courteau}}]{Dutton07a}
{Dutton}, A.~A., {van den Bosch}, F.~C., {Dekel}, A., \& {Courteau}, S. 2007,
  \apj, 654, 27

\bibitem[{{Dutton} {et~al.}(2013){Dutton}, {Treu}, {Brewer}, {Marshall},
  {Auger}, {Barnab{\`e}}, {Koo}, {Bolton}, \& {Koopmans}}]{Dutton13a}
{Dutton}, A.~A., {Treu}, T., {Brewer}, B.~J., {et~al.} 2013, \mnras, 428, 3183

\bibitem[{{Epinat} {et~al.}(2009){Epinat}, {Contini}, {Le F{\`e}vre},
  {Vergani}, {Garilli}, {Amram}, {Queyrel}, {Tasca}, \& {Tresse}}]{Epinat09a}
{Epinat}, B., {Contini}, T., {Le F{\`e}vre}, O., {et~al.} 2009, \aap, 504, 789

\bibitem[{{Epinat} {et~al.}(2012){Epinat}, {Tasca}, {Amram}, {Contini}, {Le
  F{\`e}vre}, {Queyrel}, {Vergani}, {Garilli}, {Kissler-Patig}, {Moultaka},
  {Paioro}, {Tresse}, {Bournaud}, {L{\'o}pez-Sanjuan}, \& {Perret}}]{Epinat12a}
{Epinat}, B., {Tasca}, L., {Amram}, P., {et~al.} 2012, \aap, 539, A92

\bibitem[{{Erb} {et~al.}(2003){Erb}, {Shapley}, {Steidel}, {Pettini},
  {Adelberger}, {Hunt}, {Moorwood}, \& {Cuby}}]{Erb03a}
{Erb}, D.~K., {Shapley}, A.~E., {Steidel}, C.~C., {et~al.} 2003, \apj, 591, 101

\bibitem[{{F{\"o}rster Schreiber} {et~al.}(2009){F{\"o}rster Schreiber},
  {Genzel}, {Bouch{\'e}}, {Cresci}, {Davies}, {Buschkamp}, {Shapiro},
  {Tacconi}, {Hicks}, {Genel}, {Shapley}, {Erb}, {Steidel}, {Lutz},
  {Eisenhauer}, {Gillessen}, {Sternberg}, {Renzini}, {Cimatti}, {Daddi},
  {Kurk}, {Lilly}, {Kong}, {Lehnert}, {Nesvadba}, {Verma}, {McCracken},
  {Arimoto}, {Mignoli}, \& {Onodera}}]{ForsterSchreiber09a}
{F{\"o}rster Schreiber}, N.~M., {Genzel}, R., {Bouch{\'e}}, N., {et~al.} 2009,
  \apj, 706, 1364

\bibitem[{{Genzel} {et~al.}(2017){Genzel}, {Schreiber}, {{\"U}bler}, {Lang},
  {Naab}, {Bender}, {Tacconi}, {Wisnioski}, {Wuyts}, {Alexander}, {Beifiori},
  {Belli}, {Brammer}, {Burkert}, {Carollo}, {Chan}, {Davies}, {Fossati},
  {Galametz}, {Genel}, {Gerhard}, {Lutz}, {Mendel}, {Momcheva}, {Nelson},
  {Renzini}, {Saglia}, {Sternberg}, {Tacchella}, {Tadaki}, \&
  {Wilman}}]{Genzel17a}
{Genzel}, R., {Schreiber}, N.~M.~F., {{\"U}bler}, H., {et~al.} 2017, \nat, 543,
  397

\bibitem[{{Hainline} {et~al.}(2011){Hainline}, {Blain}, {Smail}, {Alexander},
  {Armus}, {Chapman}, \& {Ivison}}]{Hainline11a}
{Hainline}, L.~J., {Blain}, A.~W., {Smail}, I., {et~al.} 2011, \apj, 740, 96

\bibitem[{{Kassin} {et~al.}(2006){Kassin}, {de Jong}, \& {Weiner}}]{Kassin06a}
{Kassin}, S.~A., {de Jong}, R.~S., \& {Weiner}, B.~J. 2006, \apj, 643, 804

\bibitem[{{Kennicutt}(1998)}]{Kennicutt98a}
{Kennicutt}, Jr., R.~C. 1998, \araa, 36, 189

\bibitem[{{Koekemoer} {et~al.}(2007){Koekemoer}, {Aussel}, {Calzetti}, {Capak},
  {Giavalisco}, {Kneib}, {Leauthaud}, {Le F{\`e}vre}, {McCracken}, {Massey},
  {Mobasher}, {Rhodes}, {Scoville}, \& {Shopbell}}]{Koekemoer07a}
{Koekemoer}, A.~M., {Aussel}, H., {Calzetti}, D., {et~al.} 2007, \apjs, 172,
  196

\bibitem[{{Korsaga} {et~al.}(2018){Korsaga}, {Carignan}, {Amram}, {Epinat}, \&
  {Jarrett}}]{Korsaga18a}
{Korsaga}, M., {Carignan}, C., {Amram}, P., {Epinat}, B., \& {Jarrett}, T.~H.
  2018, \mnras, arXiv:1804.05820

\bibitem[{{Laigle} {et~al.}(2016){Laigle}, {McCracken}, {Ilbert}, {Hsieh},
  {Davidzon}, {Capak}, {Hasinger}, {Silverman}, {Pichon}, {Coupon}, {Aussel},
  {Le Borgne}, {Caputi}, {Cassata}, {Chang}, {Civano}, {Dunlop}, {Fynbo},
  {Kartaltepe}, {Koekemoer}, {Le F{\`e}vre}, {Le Floc'h}, {Leauthaud}, {Lilly},
  {Lin}, {Marchesi}, {Milvang-Jensen}, {Salvato}, {Sanders}, {Scoville},
  {Smolcic}, {Stockmann}, {Taniguchi}, {Tasca}, {Toft}, {Vaccari}, \&
  {Zabl}}]{Laigle16a}
{Laigle}, C., {McCracken}, H.~J., {Ilbert}, O., {et~al.} 2016, \apjs, 224, 24

\bibitem[{{Lang} {et~al.}(2017){Lang}, {F{\"o}rster Schreiber}, {Genzel},
  {Wuyts}, {Wisnioski}, {Beifiori}, {Belli}, {Bender}, {Brammer}, {Burkert},
  {Chan}, {Davies}, {Fossati}, {Galametz}, {Kulkarni}, {Lutz}, {Mendel},
  {Momcheva}, {Naab}, {Nelson}, {Saglia}, {Seitz}, {Tacchella}, {Tacconi},
  {Tadaki}, {{\"U}bler}, {van Dokkum}, \& {Wilman}}]{Lang17a}
{Lang}, P., {F{\"o}rster Schreiber}, N.~M., {Genzel}, R., {et~al.} 2017, \apj,
  840, 92

\bibitem[{{Law} {et~al.}(2009){Law}, {Steidel}, {Erb}, {Larkin}, {Pettini},
  {Shapley}, \& {Wright}}]{Law09a}
{Law}, D.~R., {Steidel}, C.~C., {Erb}, D.~K., {et~al.} 2009, \apj, 697, 2057

\bibitem[{{Lovell} {et~al.}(2018){Lovell}, {Pillepich}, {Genel}, {Nelson},
  {Springel}, {Pakmor}, {Marinacci}, {Weinberger}, {Torrey}, {Vogelsberger}, \&
  {Hernquist}}]{Lovell18a}
{Lovell}, M.~R., {Pillepich}, A., {Genel}, S., {et~al.} 2018, ArXiv e-prints,
  arXiv:1801.10170

\bibitem[{{Martinsson} {et~al.}(2013{\natexlab{a}}){Martinsson}, {Verheijen},
  {Westfall}, {Bershady}, {Andersen}, \& {Swaters}}]{Martinsson13a}
{Martinsson}, T.~P.~K., {Verheijen}, M.~A.~W., {Westfall}, K.~B., {et~al.}
  2013{\natexlab{a}}, \aap, 557, A131

\bibitem[{{Martinsson} {et~al.}(2013{\natexlab{b}}){Martinsson}, {Verheijen},
  {Westfall}, {Bershady}, {Schechtman-Rook}, {Andersen}, \&
  {Swaters}}]{Martinsson13b}
---. 2013{\natexlab{b}}, \aap, 557, A130

\bibitem[{{McCracken} {et~al.}(2012){McCracken}, {Milvang-Jensen}, {Dunlop},
  {Franx}, {Fynbo}, {Le F{\`e}vre}, {Holt}, {Caputi}, {Goranova}, {Buitrago},
  {Emerson}, {Freudling}, {Hudelot}, {L{\'o}pez-Sanjuan}, {Magnard}, {Mellier},
  {M{\o}ller}, {Nilsson}, {Sutherland}, {Tasca}, \& {Zabl}}]{McCracken12a}
{McCracken}, H.~J., {Milvang-Jensen}, B., {Dunlop}, J., {et~al.} 2012, \aap,
  544, A156

\bibitem[{{McGaugh}(2016)}]{McGaugh16a}
{McGaugh}, S.~S. 2016, \apj, 816, 42

\bibitem[{{McLean} {et~al.}(2010){McLean}, {Steidel}, {Epps}, {Matthews},
  {Adkins}, {Konidaris}, {Weber}, {Aliado}, {Brims}, {Canfield}, {Cromer},
  {Fucik}, {Kulas}, {Mace}, {Magnone}, {Rodriguez}, {Wang}, \&
  {Weiss}}]{McLean10a}
{McLean}, I.~S., {Steidel}, C.~C., {Epps}, H., {et~al.} 2010, in \procspie,
  Vol. 7735, Ground-based and Airborne Instrumentation for Astronomy III,
  77351E--77351E--12

\bibitem[{{McLean} {et~al.}(2012){McLean}, {Steidel}, {Epps}, {Konidaris},
  {Matthews}, {Adkins}, {Aliado}, {Brims}, {Canfield}, {Cromer}, {Fucik},
  {Kulas}, {Mace}, {Magnone}, {Rodriguez}, {Rudie}, {Trainor}, {Wang}, {Weber},
  \& {Weiss}}]{McLean12a}
{McLean}, I.~S., {Steidel}, C.~C., {Epps}, H.~W., {et~al.} 2012, in \procspie,
  Vol. 8446, Ground-based and Airborne Instrumentation for Astronomy IV, 84460J

\bibitem[{{Pelliccia} {et~al.}(2017){Pelliccia}, {Tresse}, {Epinat}, {Ilbert},
  {Scoville}, {Amram}, {Lemaux}, \& {Zamorani}}]{Pelliccia17a}
{Pelliccia}, D., {Tresse}, L., {Epinat}, B., {et~al.} 2017, \aap, 599, A25

\bibitem[{{Peng} {et~al.}(2002){Peng}, {Ho}, {Impey}, \& {Rix}}]{Peng02a}
{Peng}, C.~Y., {Ho}, L.~C., {Impey}, C.~D., \& {Rix}, H.-W. 2002, \aj, 124, 266

\bibitem[{{Planck Collaboration} {et~al.}(2016){Planck Collaboration}, {Ade},
  {Aghanim}, {Arnaud}, {Ashdown}, {Aumont}, {Baccigalupi}, {Banday},
  {Barreiro}, {Bartlett}, \& et~al.}]{Planck-Collaboration16a}
{Planck Collaboration}, {Ade}, P.~A.~R., {Aghanim}, N., {et~al.} 2016, \aap,
  594, A13

\bibitem[{{Price} {et~al.}(2016){Price}, {Kriek}, {Shapley}, {Reddy},
  {Freeman}, {Coil}, {de Groot}, {Shivaei}, {Siana}, {Azadi}, {Barro},
  {Mobasher}, {Sanders}, \& {Zick}}]{Price16a}
{Price}, S.~H., {Kriek}, M., {Shapley}, A.~E., {et~al.} 2016, \apj, 819, 80

\bibitem[{{Rubin} {et~al.}(1965){Rubin}, {Burbidge}, {Burbidge}, {Crampin}, \&
  {Prendergast}}]{Rubin65a}
{Rubin}, V.~C., {Burbidge}, E.~M., {Burbidge}, G.~R., {Crampin}, D.~J., \&
  {Prendergast}, K.~H. 1965, \apj, 141, 759

\bibitem[{{Rubin} {et~al.}(1978){Rubin}, {Thonnard}, \& {Ford}}]{Rubin78a}
{Rubin}, V.~C., {Thonnard}, N., \& {Ford}, Jr., W.~K. 1978, \apjl, 225, L107

\bibitem[{{Salvatier} {et~al.}(2016){Salvatier}, {Wiecki}, \&
  {Fonnesbeck}}]{Salvatier16a}
{Salvatier}, J., {Wiecki}, T.~V., \& {Fonnesbeck}, C. 2016, {PyMC3: Python
  probabilistic programming framework}, Astrophysics Source Code Library,
  ascl:1610.016

\bibitem[{{Scoville} {et~al.}(2016){Scoville}, {Sheth}, {Aussel}, {Vanden
  Bout}, {Capak}, {Bongiorno}, {Casey}, {Murchikova}, {Koda},
  {{\'A}lvarez-M{\'a}rquez}, {Lee}, {Laigle}, {McCracken}, {Ilbert}, {Pope},
  {Sanders}, {Chu}, {Toft}, {Ivison}, \& {Manohar}}]{Scoville16a}
{Scoville}, N., {Sheth}, K., {Aussel}, H., {et~al.} 2016, \apj, 820, 83

\bibitem[{{S{\'e}rsic}(1963)}]{Sersic63a}
{S{\'e}rsic}, J.~L. 1963, Boletin de la Asociacion Argentina de Astronomia La
  Plata Argentina, 6, 41

\bibitem[{{Simons} {et~al.}(2017){Simons}, {Kassin}, {Weiner}, {Faber},
  {Trump}, {Heckman}, {Koo}, {Pacifici}, {Primack}, {Snyder}, \& {de la
  Vega}}]{Simons17a}
{Simons}, R.~C., {Kassin}, S.~A., {Weiner}, B.~J., {et~al.} 2017, \apj, 843, 46

\bibitem[{{Sofue} \& {Rubin}(2001)}]{Sofue01a}
{Sofue}, Y., \& {Rubin}, V.~C. 2001, \araa, 39, 137

\bibitem[{{Stott} {et~al.}(2016){Stott}, {Swinbank}, {Johnson}, {Tiley},
  {Magdis}, {Bower}, {Bunker}, {Bureau}, {Harrison}, {Jarvis}, {Sharples},
  {Smail}, {Sobral}, {Best}, \& {Cirasuolo}}]{Stott16a}
{Stott}, J.~P., {Swinbank}, A.~M., {Johnson}, H.~L., {et~al.} 2016, \mnras,
  457, 1888

\bibitem[{{Swinbank} {et~al.}(2004){Swinbank}, {Smail}, {Chapman}, {Blain},
  {Ivison}, \& {Keel}}]{Swinbank04a}
{Swinbank}, A.~M., {Smail}, I., {Chapman}, S.~C., {et~al.} 2004, \apj, 617, 64

\bibitem[{{Teklu} {et~al.}(2018){Teklu}, {Remus}, {Dolag}, {Arth}, {Burkert},
  {Obreja}, \& {Schulze}}]{Teklu18a}
{Teklu}, A.~F., {Remus}, R.-S., {Dolag}, K., {et~al.} 2018, \apjl, 854, L28

\bibitem[{{van Albada} {et~al.}(1985){van Albada}, {Bahcall}, {Begeman}, \&
  {Sancisi}}]{van-Albada85a}
{van Albada}, T.~S., {Bahcall}, J.~N., {Begeman}, K., \& {Sancisi}, R. 1985,
  \apj, 295, 305

\bibitem[{{van Starkenburg} {et~al.}(2008){van Starkenburg}, {van der Werf},
  {Franx}, {Labb{\'e}}, {Rudnick}, \& {Wuyts}}]{van-Starkenburg08a}
{van Starkenburg}, L., {van der Werf}, P.~P., {Franx}, M., {et~al.} 2008, \aap,
  488, 99

\bibitem[{{Weiner} {et~al.}(2006){Weiner}, {Willmer}, {Faber}, {Melbourne},
  {Kassin}, {Phillips}, {Harker}, {Metevier}, {Vogt}, \& {Koo}}]{Weiner06a}
{Weiner}, B.~J., {Willmer}, C.~N.~A., {Faber}, S.~M., {et~al.} 2006, \apj, 653,
  1027

\bibitem[{{White} \& {Rees}(1978)}]{White78a}
{White}, S.~D.~M., \& {Rees}, M.~J. 1978, \mnras, 183, 341

\bibitem[{{Wisnioski} {et~al.}(2015){Wisnioski}, {F{\"o}rster Schreiber},
  {Wuyts}, {Wuyts}, {Bandara}, {Wilman}, {Genzel}, {Bender}, {Davies},
  {Fossati}, {Lang}, {Mendel}, {Beifiori}, {Brammer}, {Chan}, {Fabricius},
  {Fudamoto}, {Kulkarni}, {Kurk}, {Lutz}, {Nelson}, {Momcheva}, {Rosario},
  {Saglia}, {Seitz}, {Tacconi}, \& {van Dokkum}}]{Wisnioski15a}
{Wisnioski}, E., {F{\"o}rster Schreiber}, N.~M., {Wuyts}, S., {et~al.} 2015,
  \apj, 799, 209

\bibitem[{{Wright} {et~al.}(2007){Wright}, {Larkin}, {Barczys}, {Erb},
  {Iserlohe}, {Krabbe}, {Law}, {McElwain}, {Quirrenbach}, {Steidel}, \&
  {Weiss}}]{Wright07b}
{Wright}, S.~A., {Larkin}, J.~E., {Barczys}, M., {et~al.} 2007, \apj, 658, 78

\bibitem[{{Wuyts} {et~al.}(2016){Wuyts}, {F{\"o}rster Schreiber}, {Wisnioski},
  {Genzel}, {Burkert}, {Bandara}, {Beifiori}, {Belli}, {Bender}, {Brammer},
  {Chan}, {Davies}, {Fossati}, {Galametz}, {Kulkarni}, {Lang}, {Lutz},
  {Mendel}, {Momcheva}, {Naab}, {Nelson}, {Saglia}, {Seitz}, {Tacconi},
  {Tadaki}, {{\"U}bler}, {van Dokkum}, {Wilman}, \& {Wuyts}}]{Wuyts16a}
{Wuyts}, S., {F{\"o}rster Schreiber}, N.~M., {Wisnioski}, E., {et~al.} 2016,
  \apj, 831, 149

\bibitem[{{Xue} {et~al.}(2018){Xue}, {Fu}, {Isbell}, {Ivison}, {Cooray}, \&
  {Oteo}}]{Xue18a}
{Xue}, R., {Fu}, H., {Isbell}, J., {et~al.} 2018, ArXiv e-prints,
  arXiv:1807.04291

\bibitem[{{Yuan} {et~al.}(2017){Yuan}, {Richard}, {Gupta}, {Federrath},
  {Sharma}, {Groves}, {Kewley}, {Cen}, {Birnboim}, \& {Fisher}}]{Yuan17a}
{Yuan}, T., {Richard}, J., {Gupta}, A., {et~al.} 2017, \apj, 850, 61

\end{thebibliography}
\end{document}